\documentclass[aps,prl,twocolumn,showpacs,preprintnumbers,amsmath,amssymb,superscriptaddress]{revtex4-2}
\setcitestyle{super}

\usepackage{color,soul}
\usepackage{bm}

\usepackage{graphicx}
\usepackage{dcolumn}
\usepackage{bm}

\usepackage{graphicx}
\usepackage{txfonts}
\usepackage{dcolumn}
\usepackage{bm}
\usepackage{hyperref}
\hypersetup{colorlinks,citecolor=red,filecolor=black,linkcolor=blue,urlcolor=blue}
\usepackage{lipsum}
\usepackage{subcaption}
\usepackage{subfloat}
\usepackage{caption}
\usepackage{natbib}

\usepackage[utf8]{inputenc}
\usepackage{calc}
\usepackage{accents}
\usepackage{braket}

\usepackage{stackengine}

\begin{document}
\setstcolor{red}

\title{Chiral Geometric Qubit in a Frustrated Molecular Spin Triangle}
\author{Aman Ullah}
\email{aman.ullah@uv.es}
\affiliation{Instituto de Ciencia Molecular (ICMol), Universitat de València, Paterna, Spain}
\affiliation{Department of Chemistry, University of Nevada, Reno, 1664 North Virginia Street, Reno, Nevada 89557-0216, United States
}
\author{Ziqi Hu }
\affiliation{Instituto de Ciencia Molecular (ICMol), Universitat de València, Paterna, Spain}
\author{Juan Aragó }
\affiliation{Instituto de Ciencia Molecular (ICMol), Universitat de València, Paterna, Spain}
\author{Sergey A. Varganov}
\email{svarganov@unr.edu}
\affiliation{Department of Chemistry, University of Nevada, Reno, 1664 North Virginia Street, Reno, Nevada 89557-0216, United States
}
\author{Alejandro Gaita-Ariño  }
\email{alejandro.gaita@uv.es}
\affiliation{Instituto de Ciencia Molecular (ICMol), Universitat de València, Paterna, Spain}

\begin{abstract}
Spin frustration in antiferromagnetically coupled triangles creates states distinguished by chirality, offering a molecular route to encoded quantum operations. Here, we show how scalar chirality and geometric phases arise in the ground doublet of a spin-$1/2$ triangle. A Dzyaloshinskii--Moriya (DM) interaction establishes the chirality of the states, while broken $C_3$ symmetry allows these states to acquire geometric phases. We show how to encode a qubit in states of opposite scalar chirality and opposite geometric phases and implement a geometric $Z$ gate that cancels the relative dynamical phase, making it insensitive to pulse-duration errors. As an example of such a chiral geometric qubit, we investigate a Cu$_3$-based paramagnetic molecular magnet, using broken-symmetry density functional theory and multireference spin--orbit calculations to estimate the strengths of the exchange and DM interactions. The results show that such paramagnetic molecules could serve as tunable platforms for implementing chiral geometric qubits.
\end{abstract}

\maketitle
Molecular spin qubits based on paramagnetic molecules offer a promising platform for quantum information science due to their chemically tunable environments.~\cite{gaita2019molecular,atzori2019second,wasielewski2020exploiting,moreno2021measuring, liu2021quantum, ullah2022electrical, ghosh2026quantum} Antiferromagnetic spin triangles are of particular interest because no collinear arrangement can simultaneously satisfy
antiparallel alignment of all three spin centers, rendering the system magnetically frustrated. This frustration forces
the spins into a non-collinear configuration whose low-energy manifold supports states of opposite chirality.~\cite{choi2006observation,trif2008spin,trif2010spin,schnack2010effects,choi2012coherent} For a half-integer spin triangle, the ground state is a spin doublet comprising two states of opposite chirality, which provide two-level encoding for a qubit.~\cite{jones2000geometric,aiello2022chirality}

In a molecular spin triangle of three antiferromagnetically coupled $S = 1/2$ centers, the isotropic Heisenberg exchange produces a fourfold-degenerate ground manifold. This degeneracy arises from a spin-doublet ground state (two
magnetic projections) together with two degenerate senses of spin circulation induced by frustration. The
Dzyaloshinskii--Moriya (DM) interaction cants the spins out of the coplanar arrangement and splits the manifold into
two Kramers doublets of opposite chirality.~\cite{dzyaloshinsky1958thermodynamic,fabreges2016exploring} Within each doublet, the two states are related through time-reversal symmetry and therefore remain degenerate until an external magnetic field lifts this degeneracy. In the equilateral triangle, the spatial cyclic symmetry $C_3$ is preserved, so the two states of the ground Kramers doublet transform according to different irreducible representations of $C_3$ and cannot be mixed by a uniform transverse field. Lowering the $C_3$ symmetry either by distorting the spin triangle~\cite{ferrer2012antisymmetric,ponomaryov2015spin,spielberg2015spin,kintzel2018molecular,islam2010first,liu2019electric,kintzel2021spin,le2025probing} or by applying an in-plane electric field that modulates the individual
exchange couplings~\cite{liu2019electric,kintzel2021spin,le2025probing} mixes the two representations, resulting in two chiral states with distinct phases. When the transverse magnetic field components $B_x$ and $B_y$ are transported around a closed loop in the $(B_x,B_y)$ plane, the two eigenstates of the ground doublet acquire opposite geometric
phases,~\cite{berry1984quantal,aharonov1987phase} thereby realizing a geometric chiral qubit. Because the encoded information resides in the geometry of the state rather than in its dynamical phase, the qubit is protected against dynamical noise and pulse-duration errors.~\cite{sjoqvist2012non,pachos1999non,zanardi1999holonomic}

Herein, we establish the underlying chirality and geometric-phase structure in a frustrated spin triangle, subsequently developing chiral-qubit encoding and geometric gate protocols that suppress dynamical errors. We then analyze the Cu$_3$-based paramagnetic molecule~\cite{ferrer2012antisymmetric} as a promising candidate for realizing such a qubit. Broken-symmetry density functional theory (BS-DFT) is employed to determine the magnetic exchange interactions, while multireference \textit{ab initio} calculations yield the local magnetic $g$-tensors and spin--orbit coupling.~\cite{nossa2012first} Together, these electronic structure methods establish a physically motivated estimate of the DM interaction scale. The resulting energy spectrum, spin chirality, and geometric phases identify this trinuclear copper complex as a promising platform for a geometric chiral qubit.

\begin{figure}[t]
  \centering
  \includegraphics[width=\linewidth]{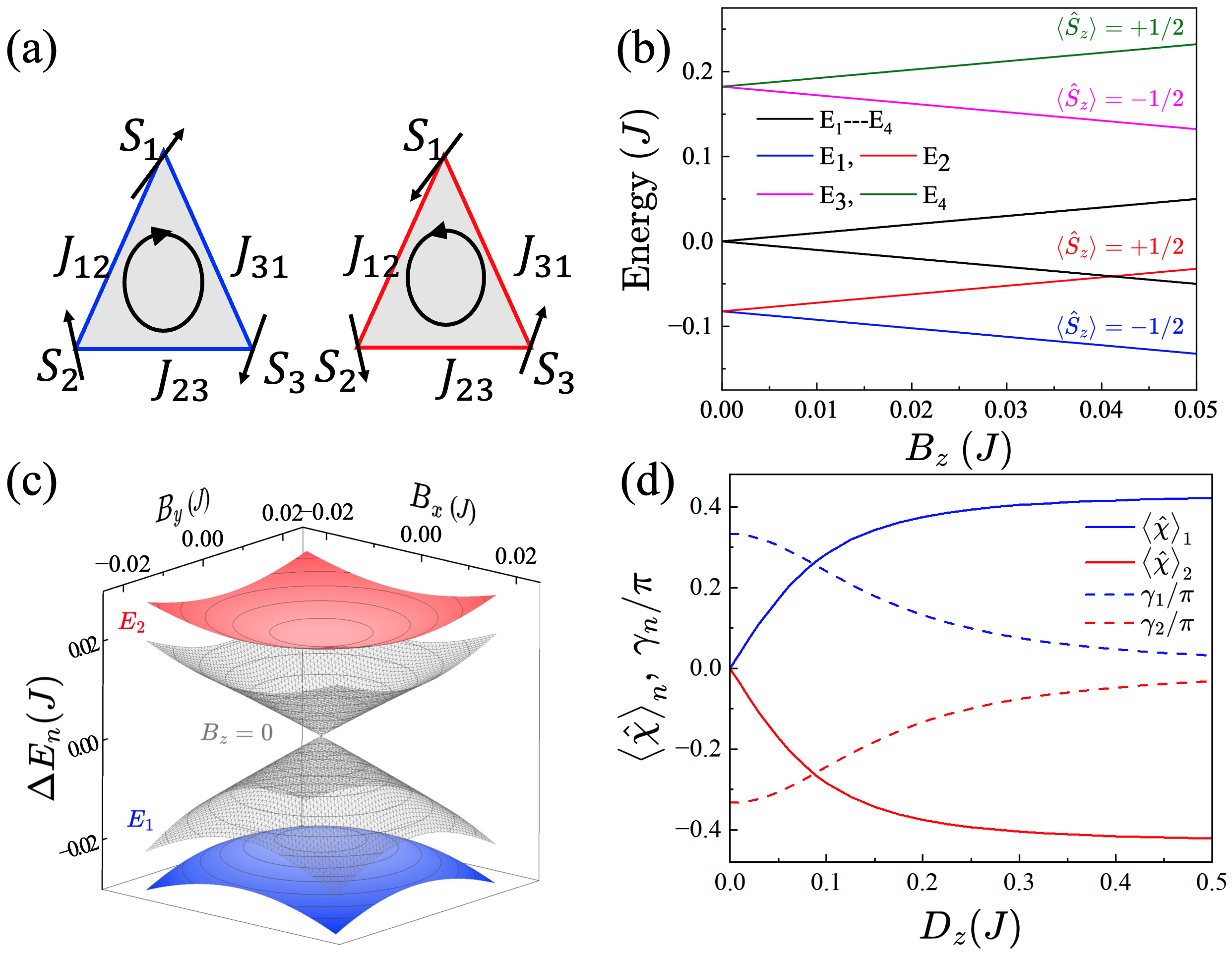}
\caption{ Energy spectrum, spin chirality, and geometric phases of the spin triangle with $J_{31}=J_w=0.80J$. (a) Schematic representation of the clockwise and counterclockwise spin alignments in the triangle. (b) Low-energy spectrum as a function of the longitudinal field $B_z$ at $D_z=0$ (black curves)
and $D_z=0.1J$ (colored curves). (c) The two lowest energy surfaces plotted against the transverse fields $B_x$ and $B_y$. Gray surfaces with a conical intersection correspond to $B_z=0$, while the red and blue surfaces correspond to
$B_z=0.02J$. (d) Scalar chirality $\langle\chi\rangle_{1,2}$ and geometric phase $\gamma_{1,2}/\pi$ as functions of the Dzyaloshinskii--Moriya interaction strength
$D_z$.}
\label{fig:1}
\end{figure}

The spin Hamiltonian of the molecular triangle contains the exchange, DM, and Zeeman terms,
\begin{equation}
\hat{H} = \sum_{\langle i,j\rangle} J_{ij}\,\hat{\mathbf{S}}_i \cdot \hat{\mathbf{S}}_j + D_z \sum_{\langle i,j\rangle}
\hat{\mathbf{z}} \cdot \left(\hat{\mathbf{S}}_i \times \hat{\mathbf{S}}_j\right) + g\,\mathbf{B} \cdot \hat{\mathbf{S}}_{\mathrm{tot}}, \quad
\hat{\mathbf{S}}_{\mathrm{tot}} = \sum_i \hat{\mathbf{S}}_i .
\label{eq:spin_hamiltonian}
\end{equation}
where $\hat{\mathbf{S}}_i$ is a spin-$1/2$ operator, $J>0$ denotes antiferromagnetic exchange, $D_z$ is the uniform out-of-plane DM coupling, $\hat{\mathbf{z}}$ is the unit vector normal to the triangle plane, $g$ is the isotropic $g$ factor, and the Bohr magneton is absorbed into the field variable $\mathbf{B}$. Both bond sums run over the three nearest-neighbor bonds of the triangle, with the bonds directed cyclically as $1 \rightarrow 2$, $2 \rightarrow 3$, and $3 \rightarrow 1$.

For three spin-$1/2$ centers, the total Hilbert space is $2 \times 2 \times 2 = 8$-dimensional. The isotropic exchange is invariant under global SU(2) spin rotations and time reversal, yielding a fourfold-degenerate ground manifold (two degenerate $S=1/2$ doublets) in the equilateral triangle. The out-of-plane DM term breaks the SU(2) symmetry down to U(1) axial symmetry around the $\hat{\mathbf{z}}$ axis. Because the DM interaction is time-reversal even, it preserves Kramers degeneracy and splits the fourfold manifold into two distinct doublets at zero field. Within each doublet, the DM term couples the spin orientation to a collective clockwise or anticlockwise spin, ensuring that the two states in a pair retain opposite chirality and spin at zero field. The Zeeman term $g\mathbf{B}\cdot\hat{\mathbf{S}}_{\mathrm{tot}}$ lifts the remaining Kramers degeneracy, making individual states addressable. A longitudinal component $B_z$ preserves the axial U(1) symmetry, whereas a transverse component $(B_x,B_y)$ also breaks this residual spin-rotation symmetry.

The spin chirality is characterized by the scalar chirality operator:
\begin{equation}
\hat{\chi}_{123} = \hat{\mathbf{S}}_1 \cdot \left(\hat{\mathbf{S}}_2 \times \hat{\mathbf{S}}_3\right),
\label{eq:scalar_chirality}
\end{equation}
which yields opposite expectation values for the two split doublets. The scalar chirality of the $n$th eigenstate $|\psi_n\rangle$ of Eq.~\eqref{eq:spin_hamiltonian}
is given by $\langle \hat{\chi}_{123} \rangle_n = \langle \psi_n | \hat{\chi}_{123} | \psi_n \rangle$.

In an equilateral triangle, the two chiral states transform as the complex-conjugate irreducible representations of $\mathrm{C}_3$ ($e^{i2\pi/3}$ and $e^{-i2\pi/3}$).~\cite{cotton1990group} Because these representations are distinct, a uniform transverse field path preserves this separation, preventing a relative geometric phase from accumulating between them along a closed transverse-field loop. Distorting the triangle breaks the $\mathrm{C}_3$ symmetry and mixes these representations; the distortion can originate from a ligand-field effect, spin-vibronic coupling, or electrostatic modulation.~\cite{ullah2025effects} As a result, the two chiral states couple and acquire a well-defined relative geometric phase under a closed transverse-field path.

Each eigenstate acquires a geometric phase when the transverse field is transported around a closed contour $C$ in the $(B_x,B_y)$ plane,~\cite{berry1984quantal,zwanziger1990berry,tomka2012geometric,zhu2006scaling}
\begin{equation}
\gamma_n = i \oint_C \left\langle \psi_n(\mathbf{B}) \middle| \nabla_{\mathbf{B}} \psi_n(\mathbf{B}) \right\rangle \cdot d\mathbf{B},
\label{eq:berry_phase}
\end{equation}
where $\nabla_{\mathbf{B}} = (\partial/\partial B_x,\partial/\partial B_y)$. This phase is evaluated numerically along a closed loop enclosing the origin, with $B_z$ held fixed, using the gauge-invariant Fukui--Hatsugai--Suzuki discretization.~\cite{fukui2005chern} The parameter space is partitioned into a discrete mesh of small squares (plaquettes). For an individual square with four ordered corner vertices $\mathbf{B}_1$, $\mathbf{B}_2$, $\mathbf{B}_3$, and $\mathbf{B}_4$ in counterclockwise order, the normalized link variable between two adjacent vertices $\mathbf{B}_j$ and $\mathbf{B}_k$ is
\begin{equation}
U_{jk}^{n} = \frac{ \left\langle \psi_n(\mathbf{B}_j) \middle| \psi_n(\mathbf{B}_k) \right\rangle}{ \left|
\left\langle \psi_n(\mathbf{B}_j) \middle| \psi_n(\mathbf{B}_k) \right\rangle\right|},
\label{eq:link_variable}
\end{equation}
The phase of each plaquette is $\Delta\gamma_n = -\arg\left( U_{12}^{n} U_{23}^{n} U_{34}^{n} U_{41}^{n}\right)$, and the total geometric phase is the sum of $\Delta\gamma_n$ over all plaquettes enclosed by the contour. Equivalently,
\begin{equation}
\gamma_n = -\arg \prod_{\langle ij\rangle \in C} U_{ij}^{n}.
\label{eq:discrete_berry_phase}
\end{equation}
where the product runs over the oriented links along the boundary of $C$.

\begin{figure*}[t]
  \centering
  \includegraphics[width=\linewidth]{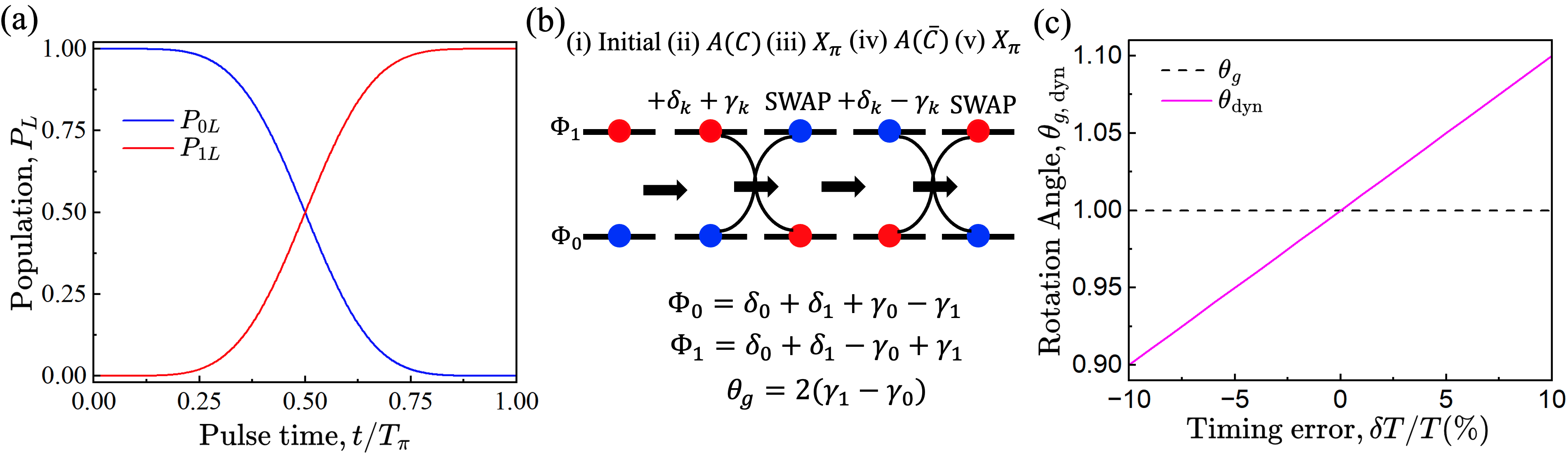}
\caption{Swap and $Z$-gate operations on the chiral geometric qubit.(a) Simulated $X_\pi$ pulse showing the complete population transfer from $|0_L\rangle$ to $|1_L\rangle$. (b) Five-step schematic of the geometric $Z$ gate starting from an initial state $|\psi\rangle=a|0_L\rangle+b|1_L\rangle$: (i) initial amplitudes, (ii) forward loop, (iii) first population swap, (iv) reversed loop, and (v) second swap restoring the basis labels. (c) Rotation angle versus fractional timing error, comparing the linear growth of the dynamical angle with the insensitivity of the geometric angle. }
\label{fig:2}

\end{figure*}

In Figure~\ref{fig:1}, we numerically demonstrate the energy spectrum, spin chirality, and geometric phases of the clockwise and counterclockwise spin configurations shown in Figure~\ref{fig:1}(a). All energy terms are scaled to the exchange coupling ($J_{ij}=1$), leaving the dimensionless ratios $D_z/J_{ij}$ and $B/J_{ij}$ as our working variables. At $D_z=0$ and $B_z=0$, the spin degrees of freedom form a fourfold-degenerate ground manifold, labeled $E_1$--$E_4$ in the black curves of Figure~\ref{fig:1}(b). Introducing a DM interaction with $D_z=0.1J$ splits this manifold into a ground and an excited Kramers doublet
($E_1$--$E_2$ and $E_3$--$E_4$, respectively), as depicted by the colored spectrum in Figure~\ref{fig:1}(b). Applying a longitudinal magnetic field further lifts this degeneracy, splitting each Kramers doublet and causing the four distinct energy levels to evolve as a function of the field.

A weak closing bond ($J_{12}=J_{23}=1$ and $J_{31}=J_w=0.80J$) breaks the underlying $\mathrm{C}_3$ symmetry. In the transverse field parameter space ($B_x,B_y\in[-0.02,0.02]$) with $B_z=0$, the ground doublet forms a conical intersection (Figure~\ref{fig:1}(c)). A finite $B_z$ lifts this degeneracy, separating the two lowest energy levels into distinct, nondegenerate surfaces. When $D_z=0$, the scalar chiralities vanish ($\langle\chi\rangle_{1,2}=0$), and the
geometric phases are given by $\gamma_{1,2}/\pi=\pm 1/3$
(Figure~\ref{fig:1}(d)). Tuning $D_z$ from $0$ to $0.5J$ is generating equal and opposite scalar chiralities while monotonically reducing the magnitude of the geometric phases. Specifically, at $D_z=0.1J$, the two states yield $\langle\chi\rangle_{1,2}=(+0.2835,-0.2835)$ and $\gamma_{1,2}/\pi=(+0.24018,-0.24419)$, confirming that both the chirality and the geometric phase are equal and opposite at a finite DM strength. As $D_z$ increases, the out-of-plane DM component dominates the transverse magnetic field, and the states
accumulate a smaller geometric phase over the same closed loop. The two phases remain approximately equal and opposite, but their magnitude decreases as the DM aligns the states more strongly along the out-of-plane axis.

The above discussion shows that the breaking of $\mathrm{C}_3$ symmetry is the crucial mechanism that allows the two chiral states to acquire a relative geometric phase, whereas the DM interaction is uniquely responsible for generating their opposing scalar chiralities.

The chiral ground doublet, $|\psi_1\rangle$ and $|\psi_2\rangle$, provides a basis for encoding a logical qubit.~\cite{scarola2004chirality,georgeot2010chirality,chiesa2023chirality,tang2026advancing} Unlike a conventional spin qubit, where the basis is defined by the spatial orientation of an individual spin (e.g., up or down), this chiral qubit stores information in the chirality of a three-spin system. We order the two eigenstates according to their scalar chirality expectation values, $\langle\psi_1|\hat{\chi}_{123}|\psi_1\rangle=+\chi_0$ and
$\langle\psi_2|\hat{\chi}_{123}|\psi_2\rangle=-\chi_0$, and define the logical basis states as
\begin{equation}
|0_L\rangle=|\psi_1\rangle,\quad |1_L\rangle=|\psi_2\rangle.
\label{eq:logical_basis}
\end{equation}

In the logical basis $\{|0_L\rangle,|1_L\rangle\}$, the physical operators have the $2\times 2$ representations
\begin{equation}
\begin{aligned} 
\hat{\chi}_{123} &= \chi_0\tau_z, \qquad \hat{S}_{\mathrm{tot}}^z = -\frac{1}{2}\tau_z,\\ \hat{S}_{\mathrm{tot}}^x &= q\tau_x, \qquad \hat{S}_{\mathrm{tot}}^y = -q\tau_y.
\end{aligned}
\label{eq:logical_operators}
\end{equation}
where $\tau_x$, $\tau_y$, and $\tau_z$ are the Pauli matrices in this basis and
$q=\langle 0_L|\hat{S}_{\mathrm{tot}}^x|1_L\rangle>0$ is the transverse transition matrix element, with the relative phase of the two basis states chosen so that $q$ is real and positive. Scalar chirality therefore acts as the logical $z$ observable, whereas the transverse-spin matrix elements provide the controls that
coherently mix the two chiral states. The projected Hamiltonian takes the effective
qubit form
\begin{equation}
\hat{H}_q=P\hat{H}P =\bar{E}I +gq(B_x\tau_x-B_y\tau_y)
-\frac{gB_z}{2}\tau_z, \label{eq:effective_qubit_hamiltonian}
\end{equation}
where $\bar{E}=(E_1+E_2)/2$ is the mean energy of the logical doublet, which contributes only a common global phase. Equation~\eqref{eq:effective_qubit_hamiltonian} shows that the longitudinal field $B_z$ tunes the logical energy splitting, while the transverse components $B_x$ and $B_y$ generate noncommuting logical $X$ and $Y$ rotations. Adiabatically
transporting these transverse controls along a closed path produces a relative geometric phase. Combining this transport with two population swaps between the two chiral states realizes a geometric logical $Z$ rotation.

We implement the population swap using a resonant, circularly polarized transverse magnetic pulse,
\begin{equation}
\begin{aligned}
B_x(t)&=Af(t)\cos(\omega_q t+\phi),\quad B_y(t)&=Af(t)\sin(\omega_q t+\phi),
\end{aligned}
\label{eq:transverse_pulse}
\end{equation}
where $\omega_q=(E_2-E_1)/\hbar$ is the qubit transition frequency, and $A$ is the peak transverse-field amplitude. The envelope $f(t)=\sin^2(\pi t/T_\pi)$ switches the pulse on and off over the duration $T_\pi$, while the carrier phase $\phi$ selects the rotation axis in the logical $xy$ plane. Starting from
the chiral ground state $|\psi(0)\rangle=|0_L\rangle$, the time-dependent
Schr\"odinger equation is solved using the Hamiltonian in Eq.~\eqref{eq:effective_qubit_hamiltonian}. The state populations at time $t$ are
$P_{0_L}(t)=|\langle 0_L|\psi(t)\rangle|^2$ and $P_{1_L}(t)=|\langle 1_L|\psi(t)\rangle|^2$. Throughout the pulse, the longitudinal field $B_z$ remains fixed. At $t=T_\pi$, the pulse completes a $\pi$ rotation, and the populations satisfy $P_{0_L}\simeq 0$ and $P_{1_L}\simeq 1$. Figure~\ref{fig:2}(a) illustrates this coherent population transfer between the two chiral basis states. Up to a global phase, the operation takes the form
\begin{equation}
\begin{aligned}
X_\pi(\phi_x)|0_L\rangle &\simeq e^{i\phi_x}|1_L\rangle,\quad
X_\pi(\phi_x)|1_L\rangle &\simeq e^{-i\phi_x}|0_L\rangle,
\end{aligned}
\label{eq:population_swap}
\end{equation}
where the swap-axis phase $\phi_x$ is determined by the carrier phase. This phase must be tracked when the pulse is used in the geometric echo sequence. The numerical simulation with parameters $A/J=0.030$, $B_z/J=0.020$, and $\hbar\omega_q/J=0.040$ yields the transfer probability $P_{1_L}(T_\pi)=0.9999999945$, corresponding to a swap error of only $5.5\times10^{-9}$.

The logical geometric $Z$ gate is implemented using the five-step echo sequence shown in Figure~\ref{fig:2}(b). The protocol consists of a forward traversal around a closed loop $C$ in the $(B_x,B_y)$ plane, a population swap, a reverse traversal around the inverted loop $\bar{C}$, and a final population swap. The two swaps exchange the logical states between traversals, so that each amplitude follows one energy branch on the way forward and the other on the way back. During a given traversal, a state $|\psi_k\rangle$ (where $k=0_L,1_L$) tracks its instantaneous energy branch and accumulates two phases:
a dynamical phase $\delta_k=-\frac{1}{\hbar}\int_0^T E_k(t)\,dt$, which depends on the instantaneous energy and the traversal time, and a geometric phase $\gamma_k$ (Eq.~\eqref{eq:berry_phase}), which depends only on the closed path $C$. The state evolves as
\begin{equation}
|\psi_k\rangle \xrightarrow{\mathcal{A}(C)} e^{i(\delta_k+\gamma_k)}|\psi_k\rangle.
\label{eq:adiabatic_evolution}
\end{equation}

Under the reverse traversal, a state $|\psi_k\rangle$ retains its dynamical phase, $\delta_k(\bar{C})=\delta_k(C)$, because the phase depends on the time integral of the energy, while the sign of its geometric phase is flipped, $\gamma_k(\bar{C})=-\gamma_k(C)$, because the path orientation is inverted.

Consequently, an amplitude starting in state $|0_L\rangle$ accumulates a total phase of $\Phi_0=\delta_0+\delta_1+\gamma_0-\gamma_1$ over the complete sequence, while an amplitude starting in $|1_L\rangle$ accumulates $\Phi_1=\delta_0+\delta_1-\gamma_0+\gamma_1$. Both amplitudes share the exact same total
dynamical contribution $(\delta_0+\delta_1)$, which factors out entirely as an unobservable common global phase. The remaining observable  relative phase is $\Phi_1-\Phi_0=2(\gamma_1-\gamma_0)$, proving that the relative dynamical phases cancel out while the geometric phase contributions add constructively. After removing the global phase factor, the sequence implements a pure logical rotation:
\begin{equation}
U_g=R_z(\theta_g)=
\begin{pmatrix}
e^{-i\theta_g/2} & 0\\
0 & e^{+i\theta_g/2}
\end{pmatrix},
\label{eq:geometric_z_gate}
\end{equation}
where $\theta_g=2(\gamma_1-\gamma_0)$ is the rotation angle about the logical $z$-axis. It is insensitive to fluctuations in pulse duration and depends only on the geometric area enclosed by the path $C$, provided the evolution remains adiabatic.

The cancellation of the relative dynamical phase distinguishes the geometric chiral qubit from a conventional spin qubit, whose $Z$ gate is entirely dynamical. In a conventional dynamical gate, the rotation angle is determined by the product of the energy splitting and the pulse duration; consequently, any deviation of the pulse duration from its intended value $T$ translates directly into a rotation-angle error. We parametrize this deviation as $T'=T+\delta T$, where $\delta T$ is the pulse-duration error. For an isolated spin-$1/2$ governed by the Hamiltonian $H_{\mathrm{spin}}=gB_z\sigma_z/2$, the two eigenstates have energies $E_\pm=\pm gB_z/2$, yielding a relative energy splitting of $\Delta E=gB_z$. A $Z$ rotation by an angle $\theta_{\mathrm{dyn}}=\Delta E T/\hbar$ is implemented by evolving the system for a duration $T$ such that
\begin{equation}
R_z(\theta_{\mathrm{dyn}}) =\exp\left(-\frac{i\theta_{\mathrm{dyn}}\sigma_z}{2} \right).
\label{eq:dynamical_z_gate}
\end{equation}

A systematic timing error $T'=T+\delta T$ therefore gives
$\theta_{\mathrm{dyn}}' =\theta_{\mathrm{dyn}}(1+\delta T/T)$, so a fractional error $\delta T/T$ in the pulse duration produces the same fractional error in the rotation angle (Figure~\ref{fig:2}(c)). For a 10\% duration error, the dynamical angle is off by 10\%.

In the geometric gate, by contrast, a fractional error of the same magnitude changes only the duration of the adiabatic traversals, while the durations of the $X_\pi$ pulses remain fixed. Since $\theta_g=2(\gamma_1-\gamma_0)$ depends only on the path, it is unchanged when the path is traversed faster or slower, provided the evolution remains adiabatic. Figure~\ref{fig:2}(c) shows that the dynamical angle $\theta_{\mathrm{dyn}}$ varies linearly with the pulse-duration error, whereas the geometric angle $\theta_g$ remains essentially constant.

\begin{figure}[t]
  \centering
  \includegraphics[width=\linewidth]{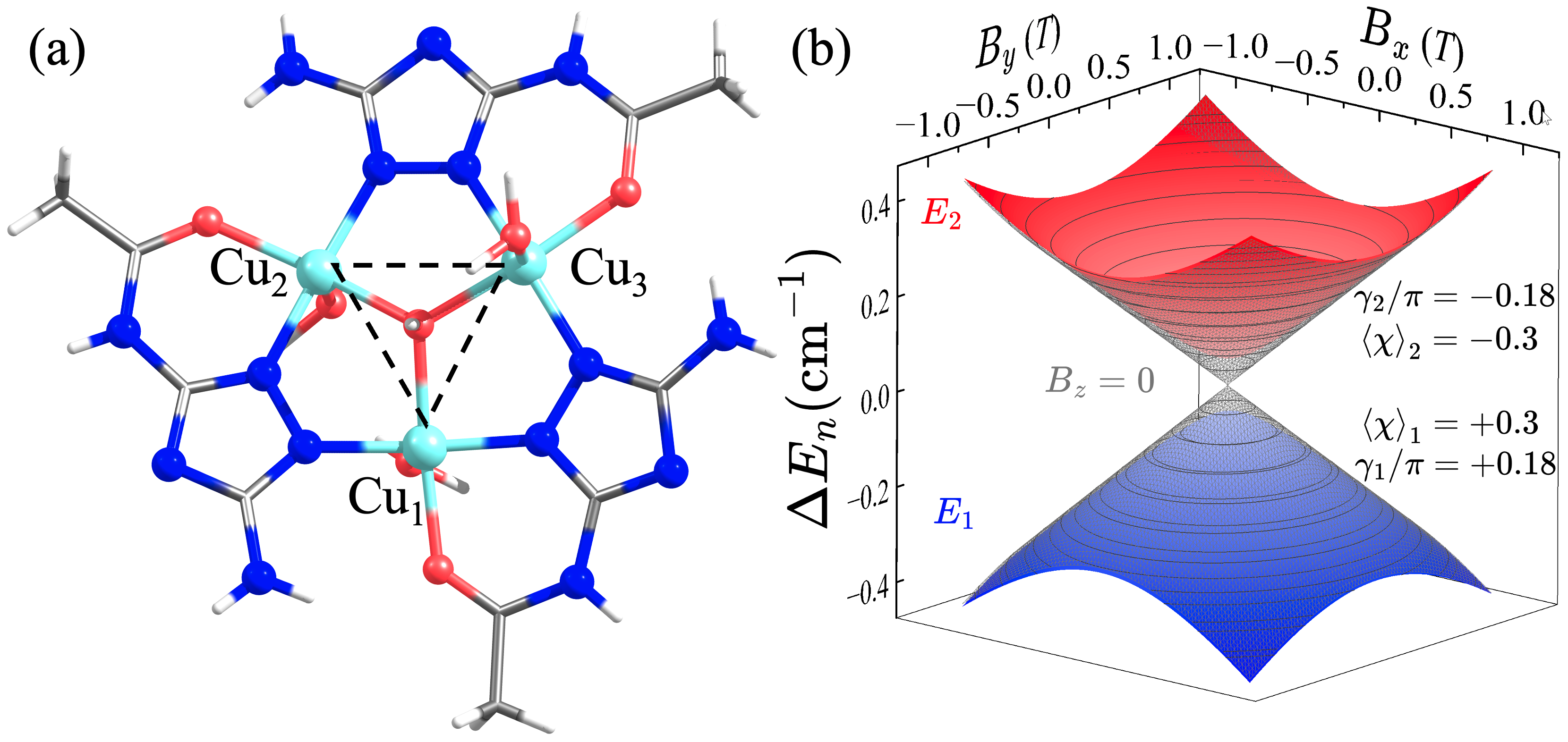}
\caption{Spin chirality and geometric phases in the Cu$_3$OH triangle. (a) Molecular structure of the trinuclear cation Cu$_3$OH. (b) Two lowest energy surfaces as functions of the transverse magnetic-field components $B_x$ and $B_y$, calculated using the effective Hamiltonian in Eq.~\eqref{eq:spin_hamiltonian}. At $B_z=0$, the Kramers ground doublet forms a conical intersection at the origin (gray surfaces). A longitudinal field $B_z=0.1$~T lifts the degeneracy and produces the separated lower ($E_1$, blue) and upper ($E_2$, red) surfaces. For the closed transverse-field loop with $B_x,B_y\in[-1.0,1.0]$~T, the two states possess
opposite scalar chiralities, $\langle\chi\rangle_{1,2}=\pm 0.30$, and accumulate opposite geometric phases,
$\gamma_{1,2}/\pi=\pm 0.18$. Energies are centered relative to the mean energy of the two states at $B_x=B_y=0$. }
\label{fig:3}
\end{figure}

We next determine the chirality and geometric phases in the Cu$_3$-based molecular triangle [Cu$_3$($\mu_3$-OH)(aaat)$_3$(H$_2$O)$_3$] (ClO$_4$)$_2\cdot$3H$_2$O, where aaat denotes the coordinating tridentate ligand.~\cite{ferrer2012antisymmetric} The cationic unit  Cu$_3$($\mu_3$-OH)(aaat)$_3$(H$_2$O)$_3$]$^{2+}$ (in short Cu$_3$OH) determines the magnetic properties of the complex (Figure~\ref{fig:3}(a)). In Cu$_3$OH, the three Cu--Cu distances are inequivalent, 3.361, 3.389, and 3.398~\AA, because of the asymmetric $\mu_3$-OH bridge with different Cu--O distances and Cu--O--Cu angles. This leads to inequivalent exchange couplings $J_{12}$, $J_{23}$, and $J_{31}$.

The isotropic exchange couplings $J_{12}$, $J_{23}$, and $J_{31}$ were determined using broken-symmetry density functional theory (BS-DFT) calculations performed with ORCA (see Methods). Each exchange coupling was modeled by retaining the two coupled Cu(II) centers and replacing the third with diamagnetic Zn(II), yielding the three fragments Cu$_1$--Cu$_2$--Zn$_3$, Cu$_1$--Zn$_2$--Cu$_3$, and Zn$_1$--Cu$_2$--Cu$_3$. The high-spin (HS) and antiparallel broken-symmetry (BS) solutions were obtained for each fragment. The exchange couplings were extracted using the spin-projected Yamaguchi expression,~\cite{soda2000ab} yielding $J_{12}=134.82$~cm$^{-1}$, $J_{23}=179.88$~cm$^{-1}$, and $J_{31}=185.08$~cm$^{-1}$. In the same Hamiltonian convention, the experimental magnetic data were described by two stronger interactions of $J=192$~cm$^{-1}$ and one weaker interaction of $j=150$~cm$^{-1}$,~\cite{ferrer2012antisymmetric} corresponding to an average of 178~cm$^{-1}$. The calculated average of 166.59~cm$^{-1}$ is somewhat smaller than the experimental estimate but reproduces both the antiferromagnetic character and the inequivalence of the exchange pathways. These calculated $J_{ij}$ values provide the bond-dependent isotropic energy scales used below to estimate the antisymmetric exchange.

The DM interaction, $\mathbf{D}_{ij}=D_{ij,z}\hat{\mathbf{z}}$, where $D_{ij,z}$ is the DM strength perpendicular to the Cu plane, was constructed from the crystallographic Cu positions $\mathbf{r}_1$, $\mathbf{r}_2$, and $\mathbf{r}_3$:
\begin{equation}
\hat{\mathbf{z}} = \frac{ (\mathbf{r}_2-\mathbf{r}_1)
\times (\mathbf{r}_3-\mathbf{r}_1) }{ \left|
(\mathbf{r}_2-\mathbf{r}_1) \times (\mathbf{r}_3 \mathbf{r}_1) \right|
}.
\label{eq:molecular_z_axis}
\end{equation}
The site ordering Cu$_1\rightarrow$Cu$_2\rightarrow$Cu$_3$ fixes the positive direction of $\hat{\mathbf{z}}$. The bonds are defined cyclically as $1\rightarrow2$, $2\rightarrow3$, and $3\rightarrow1$, with $\mathbf{D}_{ji}=-\mathbf{D}_{ij}$. Moriya's perturbative theory estimates the characteristic antisymmetric exchange to be of order $|\mathbf{D}_{ij}|\sim |J_{ij}|(|\Delta g|/g)$, where $\Delta g$ is the spin--orbit-induced change of the local $g$ factor from the free-electron value ($g_e=2.0023$).

For the Cu$_3$OH molecular triangle, the complete Cartesian $g$-tensor $\mathbf{g}_i$ of each Cu(II) center was obtained from site-resolved CASSCF/RASSI-SO/SINGLE\_ANISO calculations (see Methods). The effective $g$ factor, $g_{z,i}$, of site $i$ along the molecular $\hat{\mathbf{z}}$ axis was evaluated by projecting the complete tensor onto $\hat{\mathbf{z}}$ (Eq.~\eqref{eq:dm_estimate}). Its displacement from $g_e$ defines the dimensionless local spin--orbit response $\eta_i$ (Eq.~\eqref{eq:dm_estimate}). The exchange strength $J_{ij}$ of each bond was then combined with the spin--orbit responses of the two Cu centers connected by that bond:
\begin{equation}
\begin{aligned}
g_{z,i} &=
\sqrt{
\hat{\mathbf{z}}^{T} \mathbf{g}_i\mathbf{g}_i^{T}
\hat{\mathbf{z}}},\\
\eta_i &= \frac{g_{z,i}-g_e}{g_{z,i}},\\
|D_{ij,z}| &\approx J_{ij}\frac{\eta_i+\eta_j}{2}.
\end{aligned}
\label{eq:dm_estimate}
\end{equation}
Here, the superscript $T$ denotes the matrix transpose. This description retains the established Moriya order-of-magnitude scaling while incorporating the inequivalent exchange pathways and local magnetic anisotropies of the three Cu sites. It yields $(|D_{12,z}|,|D_{23,z}|,|D_{31,z}|) =(26.63,35.95,36.02)$~cm$^{-1}$. These values are similar to the independently fitted experimental molecular parameter $|D_z|^{\mathrm{exp}}=27$~cm$^{-1}$.

The exchange interactions $J_{ij}$, the DM interactions $D_{ij,z}$, and the local $g$-tensors were used to construct and diagonalize the spin Hamiltonian in Eq.~\eqref{eq:spin_hamiltonian}. At zero magnetic field, the eight eigenstates $E_1$--$E_8$ form two low-energy Kramers doublets and one higher-energy quartet. The ground Kramers doublet ($E_1$, $E_2$) is separated from the first excited doublet ($E_3$, $E_4$) by 66.92~cm$^{-1}$,
while the quartet ($E_5$--$E_8$) lies
283.35~cm$^{-1}$ above the ground state. A longitudinal field $B_z=1.0$~T along the molecular $\hat{\mathbf{z}}$ axis breaks the Kramers degeneracy of the ground doublet,
allowing the state-specific scalar chiralities to be evaluated using Eq.~\eqref{eq:scalar_chirality}.
The resulting values, $\langle\chi\rangle_{1,2}=(+0.3,-0.3)$, show that the two states carry equal and opposite scalar chirality. The ground pair ($E_1$, $E_2$) therefore defines the two chiral basis states.

To characterize the geometric response of this chiral manifold, the transverse magnetic-field components were transported around a counterclockwise square loop with $B_x,B_y\in[-1.0,1.0]$~T, while the longitudinal field was fixed at $B_z=1.0$~T. Figure~\ref{fig:3}(b) shows the corresponding two lowest energy surfaces as functions of $B_x$ and $B_y$. At $B_z=0$, the two surfaces form a conical intersection at the origin, as shown by the gray surfaces. A finite longitudinal field lifts this degeneracy and produces the separated lower ($E_1$, blue) and upper ($E_2$, red) surfaces. Transport around the closed transverse-field loop gives the geometric phases
$\gamma_{1,2}/\pi=(+0.18,-0.18)$, as evaluated using
Eq.~\eqref{eq:discrete_berry_phase}. Reversing the loop orientation reverses the sign of each phase, confirming their geometric origin. The coexistence of opposite scalar chirality and oppositely signed geometric phases in the two field-split states of the ground Kramers doublet establishes the essential ingredients for a chiral geometric qubit in the Cu$_3$OH complex.

In summary, a frustrated antiferromagnetic molecular spin-$1/2$ triangle possesses scalar chirality and geometric phases, allowing chiral qubit encoding. The two states of the ground Kramers doublet carry opposite scalar chiralities and geometric phases and form the qubit. A logical $Z$ gate can be implemented by combining two population swaps with the adiabatic transport of a transverse magnetic field around a closed loop. Because the rotation angle depends only on the enclosed geometric area, the dynamical phase cancels in the relative phase, making the gate insensitive to pulse-duration errors. Applying this framework to the Cu$_3$OH complex, broken-symmetry DFT and multireference \textit{ab initio} calculations yield exchange couplings and $g$-tensors that give a DM scale of the same order as the experimental estimate, confirming that the two lowest field-split states indeed exhibit the required opposite scalar chiralities and geometric phases. These results identify Cu$_3$OH as a promising molecular candidate for a geometric chiral qubit and demonstrate a quantum information encoding scheme that can be extended to other odd-membered antiferromagnetic rings.

\clearpage

\section{Theoretical Methods}

\noindent\textbf{Broken-symmetry DFT.}
Exchange interactions between the Cu(II) ions were calculated at the experimental geometry using the ORCA package (version 5.0.3).~\cite{neese2022software} For each Cu--Cu pair in the Cu$_3$ triangle, the exchange interaction was evaluated by replacing one Cu(II) with diamagnetic Zn(II); the substitutions Cu$_1$--Cu$_2$--Zn$_3$, Cu$_1$--Zn$_2$--Cu$_3$, and Zn$_1$--Cu$_2$--Cu$_3$ yield $J_{12}$, $J_{23}$, and $J_{31}$, respectively. Unrestricted Kohn--Sham calculations employed the B3LYP functional, the def2-TZVPP orbital basis on all atoms, and the RIJCOSX approximation with the def2/J auxiliary basis. Tight SCF convergence was enforced in all calculations. For each pair, the high-spin triplet ($S=1$) and the broken-symmetry ($S=0$) solutions were obtained. Exchange couplings were extracted using the Yamaguchi spin-projection formula for the Heisenberg--Dirac--van Vleck Hamiltonian~\cite{soda2000ab} and converted from the ORCA convention
($\hat{H}_{\mathrm{ex}}= -2J_{\mathrm{ORCA}}\, \hat{\mathbf{S}}_i\cdot\hat{\mathbf{S}}_j$) to that of Eq.~\eqref{eq:spin_hamiltonian} via $J_{ij}=-2J_{\mathrm{ORCA}}$, so that $J_{ij}>0$ denotes antiferromagnetic coupling.

\noindent\textbf{\textit{Ab initio} spin--orbit calculations.}
The $g$-tensor of each Cu(II) ion was obtained with OpenMolcas (version 25.10)~\cite{fdez2019openmolcas}. Three Zn-substituted variants of the Cu$_3$ complex were considered: Cu$_1$--Zn$_2$--Zn$_3$, Zn$_1$--Cu$_2$--Zn$_3$, and Zn$_1$--Zn$_2$--Cu$_3$. In each variant, two Cu(II) ions were replaced by diamagnetic Zn(II) to isolate the magnetic response of the remaining Cu(II) center. The electronic structure of each variant was calculated at the experimental geometry,~\cite{ferrer2012antisymmetric} with no prior optimization. Scalar-relativistic effects were included through the Douglas--Kroll--Hess transformation, using the ANO-RCC-VDZP basis set for all atoms, and Cholesky decomposition of the two-electron integrals was employed. The active space consisted of nine electrons in the five $3d$ orbitals of the Cu(II) ion. The molecular orbitals were optimized at the CASSCF level with state averaging over five doublet states, using equal weights. The resulting wave functions were then mixed by spin--orbit coupling through the RASSI approach, and the combined effect of the ligand field and the spin--orbit coupling was analyzed with SINGLE\_ANISO~\cite{ungur2017ab,chibotaru2012ab} to obtain the $g$-tensor and principal magnetic axes of the ground Kramers doublet.

These local $g$-tensors, together with the BS-DFT exchange couplings and the magnetic dipole--dipole interaction evaluated from the experimental geometry, were then combined in POLY\_ANISO~\cite{ungur2011magnetic} to obtain the four molecular Kramers doublets. The SINGLE\_ANISO output was used directly as the input to POLY\_ANISO, ensuring that the local $g$-tensor frames of reference were transferred consistently across the three sites. For three spin-$1/2$ centers, POLY\_ANISO combined the ground Kramers doublet from each of the three single Cu ion calculations in the $2\times2\times2=8$-dimensional product space.

\section{ASSOCIATED CONTENT}
\noindent{\textbf{Supporting Information}}\\
Additional details of the spin Hamiltonian and its
matrix representation; energy spectra, scalar chirality,
and geometric phases under different combinations of
exchange, DM, and Zeeman interactions;
SINGLE\_ANISO $g$-tensors; POLY\_ANISO energy levels
and $g$-tensors; and Cartesian coordinates of the
experimental geometry.

\section{AUTHOR INFORMATION}
\noindent\textbf{ORCID}\\
Aman Ullah: 0000-0001-5050-845X\\
Ziqi Hu: 0000-0003-1830-1584\\
Juan Aragó: 0000-0002-0415-9946\\
Sergey A. Varganov: 0000-0001-8301-3891\\
Alejandro Gaita-Ariño: 0000-0002-1600-8627 

\section{Notes}
The authors declare no competing financial interests.

\section{ACKNOWLEDGEMENTS}
This material is based upon work supported by the U.S. Department of Energy, Office of Science, Office of Basic Energy Sciences Established Program to Stimulate Competitive Research under Award Number DE-SC0022178. 

\bibliographystyle{apsrev4-2}
\bibliography{ref}

@article{gaita2019molecular,
  title={Molecular spins for quantum computation},
  author={Gaita-Ari{\~n}o, Alejandro and Luis, Fernando and Hill, Stephen and Coronado, Eugenio},
  journal={Nature chemistry},
  volume={11},
  number={4},
  pages={301--309},
  year={2019},
  publisher={Nature Publishing Group UK London}
}

@article{atzori2019second,
  title={The second quantum revolution: role and challenges of molecular chemistry},
  author={Atzori, Matteo and Sessoli, Roberta},
  journal={Journal of the American Chemical Society},
  volume={141},
  number={29},
  pages={11339--11352},
  year={2019},
  publisher={ACS Publications}
}

@article{wasielewski2020exploiting,
  title={Exploiting chemistry and molecular systems for quantum information science},
  author={Wasielewski, Michael R and Forbes, Malcolm DE and Frank, Natia L and Kowalski, Karol and Scholes, Gregory D and Yuen-Zhou, Joel and Baldo, Marc A and Freedman, Danna E and Goldsmith, Randall H and Goodson III, Theodore and others},
  journal={Nature Reviews Chemistry},
  volume={4},
  number={9},
  pages={490--504},
  year={2020},
  publisher={Nature Publishing Group UK London}
}

@article{moreno2021measuring,
  title={Measuring molecular magnets for quantum technologies},
  author={Moreno-Pineda, Eufemio and Wernsdorfer, Wolfgang},
  journal={Nature Reviews Physics},
  volume={3},
  number={9},
  pages={645--659},
  year={2021},
  publisher={Nature Publishing Group UK London}
}

@article{choi2006observation,
  title={Observation of a half step magnetization in the $\{$Cu 3$\}$-type triangular spin ring},
  author={Choi, Kwang-Yong and Matsuda, Yasuhiro H and Nojiri, Hiroyuki and Kortz, U and Hussain, F and Stowe, Ashley C and Ramsey, Chris and Dalal, Naresh S},
  journal={Physical review letters},
  volume={96},
  number={10},
  pages={107202},
  year={2006},
  publisher={APS}
}

@article{trif2008spin,
  title={Spin-electric coupling in molecular magnets},
  author={Trif, Mircea and Troiani, Filippo and Stepanenko, Dimitrije and Loss, Daniel},
  journal={Physical review letters},
  volume={101},
  number={21},
  pages={217201},
  year={2008},
  publisher={APS}
}

@article{trif2010spin,
  title={Spin electric effects in molecular antiferromagnets},
  author={Trif, Mircea and Troiani, Filippo and Stepanenko, Dimitrije and Loss, Daniel},
  journal={Physical Review B—Condensed Matter and Materials Physics},
  volume={82},
  number={4},
  pages={045429},
  year={2010},
  publisher={APS}
}

@article{schnack2010effects,
  title={Effects of frustration on magnetic molecules: a survey from Olivier Kahn until today},
  author={Schnack, J{\"u}rgen},
  journal={Dalton Transactions},
  volume={39},
  number={20},
  pages={4677--4686},
  year={2010},
  publisher={The Royal Society of Chemistry}
}

@article{choi2012coherent,
  title={Coherent manipulation of electron spins in the $\{$Cu 3$\}$ spin triangle complex impregnated in nanoporous silicon},
  author={Choi, K-Y and Wang, Z and Nojiri, H and Van Tol, J and Kumar, P and Lemmens, P and Bassil, BS and Kortz, U and Dalal, NS},
  journal={Physical review letters},
  volume={108},
  number={6},
  pages={067206},
  year={2012},
  publisher={APS}
}

@article{jones2000geometric,
  title={Geometric quantum computation using nuclear magnetic resonance},
  author={Jones, Jonathan A and Vedral, Vlatko and Ekert, Artur and Castagnoli, Giuseppe},
  journal={Nature},
  volume={403},
  number={6772},
  pages={869--871},
  year={2000},
  publisher={Nature Publishing Group UK London}
}

@article{aiello2022chirality,
  title={A chirality-based quantum leap},
  author={Aiello, Clarice D and Abendroth, John M and Abbas, Muneer and Afanasev, Andrei and Agarwal, Shivang and Banerjee, Amartya S and Beratan, David N and Belling, Jason N and Berche, Bertrand and Botana, Antia and others},
  journal={ACS nano},
  volume={16},
  number={4},
  pages={4989},
  year={2022}
}

@article{dzyaloshinsky1958thermodynamic,
  title={A thermodynamic theory of “weak” ferromagnetism of antiferromagnetics},
  author={Dzyaloshinsky, Igor},
  journal={Journal of physics and chemistry of solids},
  volume={4},
  number={4},
  pages={241--255},
  year={1958},
  publisher={Elsevier}
}

@article{fabreges2016exploring,
  title={Exploring metamagnetism of single crystalline EuNiGe 3 by neutron scattering},
  author={Fabr{\`e}ges, Xavier and Gukasov, Arsen and Bonville, Pierre and Maurya, A and Thamizhavel, A and Dhar, SK},
  journal={Physical Review B},
  volume={93},
  number={21},
  pages={214414},
  year={2016},
  publisher={APS}
}

@article{ferrer2012antisymmetric,
  title={Antisymmetric exchange in triangular tricopper (II) complexes: correlation among structural, magnetic, and electron paramagnetic resonance parameters},
  author={Ferrer, Sacramento and Lloret, Francesc and Pardo, Emilio and Clemente-Juan, Juan Modesto and Liu-Gonzalez, Malva and Garc{\'\i}a-Granda, Santiago},
  journal={Inorganic chemistry},
  volume={51},
  number={2},
  pages={985--1001},
  year={2012},
  publisher={ACS Publications}
}

@article{ponomaryov2015spin,
  title={Spin decoherence processes in the S= 1/2 scalene triangular cluster (Cu3 (OH))},
  author={Ponomaryov, Alexey N and Kim, Namseok and Jang, Zee Hoon and Tol, J van and Koo, Hyun-Joo and Law, Joseph Matthew and Suh, Byoung Jin and Yoon, Sungho and Choi, Kwang Yong},
  journal={New Journal of Physics},
  volume={17},
  number={3},
  pages={033042},
  year={2015},
  publisher={IOP Publishing}
}

@article{spielberg2015spin,
  title={A spin-frustrated trinuclear copper complex based on triaminoguanidine with an energetically well-separated degenerate ground state},
  author={Spielberg, Eike T and Gilb, Aksana and Plaul, Daniel and Geibig, Daniel and Hornig, David and Schuch, Dirk and Buchholz, Axel and Ardavan, Arzhang and Plass, Winfried},
  journal={Inorganic Chemistry},
  volume={54},
  number={7},
  pages={3432--3438},
  year={2015},
  publisher={ACS Publications}
}

@article{kintzel2018molecular,
  title={Molecular electronic spin qubits from a spin-frustrated trinuclear copper complex},
  author={Kintzel, Benjamin and B{\"o}hme, Michael and Liu, Junjie and Burkhardt, Anja and Mrozek, Jakub and Buchholz, Axel and Ardavan, Arzhang and Plass, Winfried},
  journal={Chemical communications},
  volume={54},
  number={92},
  pages={12934--12937},
  year={2018},
  publisher={The Royal Society of Chemistry}
}

@article{islam2010first,
  title={First-principles study of spin-electric coupling in a $\{$Cu 3$\}$ single molecular magnet},
  author={Islam, M Fhokrul and Nossa, Javier F and Canali, Carlo M and Pederson, Mark},
  journal={Physical Review B—Condensed Matter and Materials Physics},
  volume={82},
  number={15},
  pages={155446},
  year={2010},
  publisher={APS}
}

@article{liu2019electric,
  title={Electric field control of spins in molecular magnets},
  author={Liu, Junjie and Mrozek, Jakub and Myers, William K and Timco, Grigore A and Winpenny, Richard EP and Kintzel, Benjamin and Plass, Winfried and Ardavan, Arzhang},
  journal={Physical review letters},
  volume={122},
  number={3},
  pages={037202},
  year={2019},
  publisher={APS}
}

@article{kintzel2021spin,
  title={Spin--electric coupling in a cobalt (ii)-based spin triangle revealed by electric-field-modulated electron spin resonance spectroscopy},
  author={Kintzel, Benjamin and Fittipaldi, Maria and B{\"o}hme, Michael and Cini, Alberto and Tesi, Lorenzo and Buchholz, Axel and Sessoli, Roberta and Plass, Winfried},
  journal={Angewandte Chemie International Edition},
  volume={60},
  number={16},
  pages={8832--8838},
  year={2021},
  publisher={Wiley Online Library}
}

@article{le2025probing,
  title={Probing spin-electric transitions in a molecular exchange qubit},
  author={Le Mardel{\'e}, Florian and Mohelsk{\`y}, Ivan and Wyzula, Jan and Orlita, Milan and Turek, Philippe and Troiani, Filippo and Boudalis, Athanassios K},
  journal={nature communications},
  volume={16},
  number={1},
  pages={1198},
  year={2025},
  publisher={Nature Publishing Group UK London}
}

@article{aharonov1987phase,
  title={Phase change during a cyclic quantum evolution},
  author={Aharonov, Yakir and Anandan, Jeeva},
  journal={Physical Review Letters},
  volume={58},
  number={16},
  pages={1593},
  year={1987},
  publisher={APS}
}

@article{sjoqvist2012non,
  title={Non-adiabatic holonomic quantum computation},
  author={Sj{\"o}qvist, Erik and Tong, Dian-Min and Mauritz Andersson, L and Hessmo, Bj{\"o}rn and Johansson, Markus and Singh, Kuldip},
  journal={New Journal of Physics},
  volume={14},
  number={10},
  pages={103035},
  year={2012},
  publisher={IOP Publishing}
}

@article{pachos1999non,
  title={Non-Abelian Berry connections for quantum computation},
  author={Pachos, Jiannis and Zanardi, Paolo and Rasetti, Mario},
  journal={Physical Review A},
  volume={61},
  number={1},
  pages={010305},
  year={1999},
  publisher={APS}
}

@article{zanardi1999holonomic,
  title={Holonomic quantum computation},
  author={Zanardi, Paolo and Rasetti, Mario},
  journal={Physics Letters A},
  volume={264},
  number={2-3},
  pages={94--99},
  year={1999},
  publisher={Elsevier}
}

@article{nossa2012first,
  title={First-principles studies of spin-orbit and Dzyaloshinskii-Moriya interactions in the $\{$Cu 3$\}$ single-molecule magnet},
  author={Nossa, JF and Islam, MF and Canali, Carlo M and Pederson, MR},
  journal={Physical Review B—Condensed Matter and Materials Physics},
  volume={85},
  number={8},
  pages={085427},
  year={2012},
  publisher={APS}
}

@book{cotton1990group,
  title={Group Theory},
  author={Cotton, F Albert},
  year={1990},
  publisher={Wiley, New York}
}

@article{ullah2025effects,
  title={Effects of Coupling Between Chiral Vibrations and Spins in Molecular Magnets},
  author={Ullah, Aman and Varganov, Sergey A and Barlas, Yafis},
  journal={arXiv preprint arXiv:2505.10904},
  year={2025}
}

@article{berry1984quantal,
  title={Quantal phase factors accompanying adiabatic changes},
  author={Berry, Michael Victor},
  journal={Proceedings of the Royal Society of London. A. Mathematical and Physical Sciences},
  volume={392},
  number={1802},
  pages={45--57},
  year={1984},
  publisher={The Royal Society London}
}

@article{zwanziger1990berry,
  title={Berry's phase},
  author={Zwanziger, Josef W and Koenig, Marianne and Pines, Alex},
  journal={Annual review of physical chemistry},
  volume={41},
  year={1990}
}

@article{tomka2012geometric,
  title={Geometric phase contribution to quantum nonequilibrium many-body dynamics},
  author={Tomka, Michael and Polkovnikov, Anatoli and Gritsev, Vladimir},
  journal={Physical review letters},
  volume={108},
  number={8},
  pages={080404},
  year={2012},
  publisher={APS}
}

@article{zhu2006scaling,
  title={Scaling of geometric phases close to the quantum phase transition in the XY spin chain},
  author={Zhu, Shi-Liang},
  journal={Physical review letters},
  volume={96},
  number={7},
  pages={077206},
  year={2006},
  publisher={APS}
}

@article{fukui2005chern,
  title={Chern numbers in discretized Brillouin zone: Efficient method of computing (spin) Hall conductances},
  author={Fukui, Takahiro and Hatsugai, Yasuhiro and Suzuki, Hiroshi},
  journal={Journal of the Physical Society of Japan},
  volume={74},
  number={6},
  pages={1674--1677},
  year={2005},
  publisher={The Physical Society of Japan}
}

@article{scarola2004chirality,
  title={Chirality in quantum computation with spin cluster qubits},
  author={Scarola, VW and Park, K and Sarma, S Das},
  journal={Physical review letters},
  volume={93},
  number={12},
  pages={120503},
  year={2004},
  publisher={APS}
}

@article{georgeot2010chirality,
  title={Chirality of triangular antiferromagnetic clusters as a qubit},
  author={Georgeot, Bertrand and Mila, F},
  journal={Physical review letters},
  volume={104},
  number={20},
  pages={200502},
  year={2010},
  publisher={APS}
}

@article{chiesa2023chirality,
  title={Chirality-induced spin selectivity: an enabling technology for quantum applications},
  author={Chiesa, Alessandro and Privitera, Alberto and Macaluso, Emilio and Mannini, Matteo and Bittl, Robert and Naaman, Ron and Wasielewski, Michael R and Sessoli, Roberta and Carretta, Stefano},
  journal={Advanced Materials},
  volume={35},
  number={28},
  pages={2300472},
  year={2023},
  publisher={Wiley Online Library}
}

@article{tang2026advancing,
  title={Advancing Room-Temperature Spin Qubits with Naphthalene Diimide-Based Chiral Covalent Organic Frameworks},
  author={Tang, Chang and Sun, Zhecheng and Wei, Huijie and Sun, Lei and Jin, Zexin},
  journal={Journal of the American Chemical Society},
  volume={148},
  number={10},
  pages={10684--10691},
  year={2026},
  publisher={ACS Publications}
}

@article{soda2000ab,
  title={Ab initio computations of effective exchange integrals for H--H, H--He--H and Mn2O2 complex: comparison of broken-symmetry approaches},
  author={Soda, T and Kitagawa, Y and Onishi, T and Takano, Y and Shigeta, Y and Nagao, H and Yoshioka, Y and Yamaguchi, K},
  journal={Chemical Physics Letters},
  volume={319},
  number={3-4},
  pages={223--230},
  year={2000},
  publisher={Elsevier}
}

@article{neese2022software,
  title={Software update: The ORCA program system—Version 5.0},
  author={Neese, Frank},
  journal={Wiley Interdisciplinary Reviews: Computational Molecular Science},
  volume={12},
  number={5},
  pages={e1606},
  year={2022},
  publisher={Wiley Online Library}
}

@article{fdez2019openmolcas,
  title={OpenMolcas: From source code to insight},
  author={Fdez. Galv{\'a}n, Ignacio and Vacher, Morgane and Alavi, Ali and Angeli, Celestino and Aquilante, Francesco and Autschbach, Jochen and Bao, Jie J and Bokarev, Sergey I and Bogdanov, Nikolay A and Carlson, Rebecca K and others},
  journal={Journal of chemical theory and computation},
  volume={15},
  number={11},
  pages={5925--5964},
  year={2019},
  publisher={ACS Publications}
}

@article{ungur2017ab,
  title={Ab initio crystal field for lanthanides},
  author={Ungur, Liviu and Chibotaru, Liviu F},
  journal={Chemistry--A European Journal},
  volume={23},
  number={15},
  pages={3708--3718},
  year={2017},
  publisher={Wiley Online Library}
}

@article{chibotaru2012ab,
  title={Ab initio calculation of anisotropic magnetic properties of complexes. I. Unique definition of pseudospin Hamiltonians and their derivation},
  author={Chibotaru, Liviu F and Ungur, Liviu},
  journal={The Journal of chemical physics},
  volume={137},
  number={6},
  year={2012},
  publisher={AIP Publishing}
}

@article{ungur2011magnetic,
  title={Magnetic anisotropy in the excited states of low symmetry lanthanide complexes},
  author={Ungur, Liviu and Chibotaru, Liviu F},
  journal={Physical Chemistry Chemical Physics},
  volume={13},
  number={45},
  pages={20086--20090},
  year={2011},
  publisher={The Royal Society of Chemistry}
}

@article{ullah2022electrical,
  title={Electrical two-qubit gates within a pair of clock-qubit magnetic molecules},
  author={Ullah, Aman and Hu, Ziqi and Cerd{\'a}, Jes{\'u}s and Arag{\'o}, Juan and Gaita-Ari{\~n}o, Alejandro},
  journal={npj Quantum Information},
  volume={8},
  number={1},
  pages={133},
  year={2022},
  publisher={Nature Publishing Group UK London}
}

@article{ghosh2026quantum,
  title={Quantum coherence enhancement through control of metal--ligand covalency: modulating spin--orbit coupling in isostructural molecular qubits},
  author={Ghosh, Subrata and Oyala, Paul H and Fizer, Maksym and Dergachev, Vsevolod D and Varganov, Sergey A and Frank, Natia L},
  journal={Chemical Science},
  volume={17},
  number={20},
  pages={10029--10046},
  year={2026},
  publisher={The Royal Society of Chemistry}
}

@article{liu2021quantum,
  title={Quantum coherent spin--electric control in a molecular nanomagnet at clock transitions},
  author={Liu, Junjie and Mrozek, Jakub and Ullah, Aman and Duan, Yan and Baldov{\'\i}, Jos{\'e} J and Coronado, Eugenio and Gaita-Arino, Alejandro and Ardavan, Arzhang},
  journal={Nature Physics},
  volume={17},
  number={11},
  pages={1205--1209},
  year={2021},
  publisher={Nature Publishing Group UK, London}
}
\end{document}